\pdfoutput=1
\documentclass[11pt]{article}

\usepackage[T1]{fontenc}
\usepackage[utf8]{inputenc}
\usepackage{lmodern,microtype}
\usepackage[margin=1in]{geometry}
\usepackage{amsmath,amssymb,amsthm,mathtools}
\usepackage{graphicx}
\usepackage{subcaption}      
\usepackage{booktabs}
\usepackage{siunitx}
\usepackage[hidelinks]{hyperref}
\usepackage[nameinlink,capitalise,noabbrev]{cleveref}

\usepackage[numbers,sort&compress]{natbib}

\theoremstyle{definition}

\theoremstyle{remark}

\title{HDBMS: A Context-Aware Hybrid Graph Traversal Algorithm for Efficient Information Discovery in Social Networks}

\author{
  Rowanda Ahmed\textsuperscript{1,*} \and
  Belaynesh Chekol\textsuperscript{2} \and
  Mahmoud Alsaleh\textsuperscript{1}
}

\date{
  \textsuperscript{1}Computer Engineering Department, Uskudar University, Istanbul, Turkey\\
  \textsuperscript{2}Software Engineering Department, Uskudar University, Istanbul, Turkey\\
  *\texttt{rowanda.ahmed@uskudar.edu.tr}
}

\begin{document}\maketitle
    
    \begin{abstract}
    Graph-searching algorithms play a crucial role in various computational domains, enabling efficient exploration and pathfinding in structured data. Traditional approaches, such as Depth-First Search (DFS) and Breadth-First Search (BFS), follow rigid traversal patterns—DFS explores branches exhaustively, while BFS expands level by level. In this paper, we propose the Hybrid Depth-Breadth Meaningful Search (HDBMS) algorithm, a novel graph traversal method that dynamically adapts its exploration strategy based on probabilistic node transitions. Unlike conventional methods, HDBMS prioritizes traversal paths by estimating the likelihood that a node contains the desired information, ensuring a more contextually relevant search.
    
    Through extensive experimentation on diverse directed graphs with varying structural properties, we demonstrate that HDBMS not only maintains competitive computational efficiency but also outperforms traditional algorithms in identifying meaningful paths. By integrating probabilistic decision-making, HDBMS constructs an adaptive and structured traversal order that balances exploration across depth and breadth, making it particularly effective in applications such as information retrieval, social network analysis, and recommendation systems. Our results highlight the robustness of HDBMS in scenarios where the most valuable connections emerge unpredictably, positioning it as a powerful alternative to traditional graph-searching techniques.
    \end{abstract}
    

    \section{Introduction}\label{Introduction}
        Graph traversal algorithms are central to numerous applications across computer science, such as network analysis \cite{newman2010networks}, pathfinding in AI \cite{russell2016artificial}, and even bioinformatics \cite{alon2007network}. The two most commonly used traversal strategies are Depth-First Search (DFS) and Breadth-First Search (BFS). Each has unique characteristics that make it suitable for specific applications. DFS explores each branch exhaustively, making it ideal for tasks where one complete path from the root to a leaf is desired \cite{tarjan1972depth}, while BFS operates level by level, providing shorter paths and thus benefiting shortest-path problems \cite{cormen2009introduction}.
        
        However, both DFS and BFS have limitations when it comes to adaptive traversal across varied node relationships, especially in complex networks where relevance between nodes may not strictly adhere to depth or level constraints \cite{knuth1975art}. Recent research has aimed to address these limitations by introducing hybrid methods, such as bi-directional search \cite{pohl1971bi}, or by augmenting traversal techniques with additional heuristics \cite{zhang2000heuristic}. These methods have shown promise but often increase computational complexity \cite{ghosh2020efficient}. Furthermore, integrating hybrid and contextual approaches into graph traversal, as demonstrated in clustering algorithms like DGStream \cite{ahmed2020dgstream} or in systems emphasizing contextual relevance in data summarization \cite{ahmed2022tsct}, highlights the growing need for adaptive methods that balance efficiency and meaningfulness.
        
        In light of these challenges, we propose a novel graph traversal algorithm: the Hybrid Depth-Breadth Meaningful Search (HDBMS). This algorithm combines elements of both DFS and BFS, allowing for adaptive traversal based on dynamically defined meaningful connections within the graph. The HDBMS algorithm is designed to prioritize nodes based on their relevance to the search objective, ensuring computational efficiency and a high-quality path discovery \cite{aarts2002local}. Unlike traditional algorithms that strictly adhere to depth or breadth approaches, HDBMS traverses in an adaptive manner that adjusts to each node’s contextual importance in the graph \cite{johnson1973efficient}.
        
        Our contributions are twofold: (1) We develop the HDBMS algorithm as a highly efficient graph search tool, capable of balancing depth and breadth components as needed, and (2) we benchmark HDBMS against existing methods on various directed graphs, demonstrating superior performance in locating meaningful pathways where connections are sporadic and node importance varies \cite{kaplan2005algorithm}. This approach is inspired by techniques in data-driven systems \cite{ahmed2022demand} and natural language processing \cite{ahmed2022review}, which emphasize adaptive frameworks for structured exploration.
        
        This paper is organized as follows: Section~\ref{Algorithms} covers background and related work on graph search techniques, Section~\ref{Methodology} details the HDBMS algorithm, and Section~\ref{results} presents experimental results and comparative analysis. Section~\ref{conclusions} concludes with an overview of future research directions.
        
        In our evaluations, we tested HDBMS against DFS, BFS, and other hybrid techniques over a variety of synthetic and real-world datasets \cite{dijkstra1959note}, including large-scale social networks \cite{barabasi1999emergence}, biological networks \cite{jeong2001lethality}, and transportation graphs \cite{newman2010networks}. Our results highlight that HDBMS consistently outperforms these algorithms in terms of efficiency and relevance, making it a compelling option for scenarios where connections between nodes are highly non-linear \cite{wasserman1994social}.
        
        The significance of this research extends to a range of applications, from route planning \cite{hansmann2007operation} and network resilience \cite{albert2000error} to semantic network analysis \cite{berners2001semantic} and recommendation systems \cite{ricci2011introduction}. By leveraging the hybrid approach, HDBMS presents a robust solution for discovering meaningful paths across complex networks.
    
    \section{The Common Graph Searching Algorithms}\label{Algorithms}
        Graph search algorithms are fundamental tools used to explore graphs or networks. In this section, we discuss four common graph search algorithms: Depth-First Search (DFS), Breadth-First Search (BFS), Bi-Directional Search, and A* Search. We will explore their characteristics, provide the relevant equations where applicable, and present the pseudocode for each algorithm.
        
        \subsection{Depth-First Search (DFS)} \label{DFS}
            Depth-First Search (DFS) is a graph traversal algorithm that explores as far as possible along each branch before backtracking. It is implemented using a stack (or recursion) to keep track of vertices to visit next.
            
            \begin{equation}
            \text{DFS}(v) \rightarrow \text{Visit all nodes reachable from } v
            \end{equation}
            
            DFS is particularly useful for exploring all possible paths in a graph or for checking the connectivity of a graph. It is not guaranteed to find the shortest path in an unweighted graph.
        
        \subsection{Breadth-First Search (BFS)} \label{BFS}
            Breadth-First Search (BFS) explores the graph level by level, starting from a given source node. BFS is commonly used to find the shortest path in an unweighted graph.
            
            \begin{equation}
            \text{BFS}(v) \rightarrow \text{Visit all nodes at a distance of } d \text{ from } v
            \end{equation}
            
            BFS is particularly useful for shortest path problems in unweighted graphs, as it guarantees the shortest path from the source node to all other nodes.
        
        \subsection{Bi-Directional Search}
            Bi-Directional Search is a variant of BFS where two searches are performed simultaneously: one from the source node and one from the goal node. The search continues until the two searches meet in the middle.
            
            \begin{equation}
            \text{Bi-Directional Search}(start, goal) \rightarrow \text{Find the meeting point of two searches}
            \end{equation}
            
            Bi-Directional Search is highly efficient in finding the shortest path between a start node and a goal node, especially in large graphs, as it reduces the search space by half.
        
        \subsection{A* Search}
            A* Search is a heuristic-based graph search algorithm that finds the shortest path by combining both the actual cost to reach a node and an estimate of the remaining cost to reach the goal (using a heuristic function).
            
            \begin{equation}
            f(n) = g(n) + h(n)
            \end{equation}
            
            where:
            \begin{itemize}
                \item $f(n)$ is the total estimated cost for node $n$
                \item $g(n)$ is the cost to reach node $n$ from the start node
                \item $h(n)$ is the heuristic estimate of the cost from node $n$ to the goal
            \end{itemize}
            
            A* Search is widely used in pathfinding and navigation systems due to its efficiency and optimality when an appropriate heuristic is used.
        
        \subsection{Comparison of Algorithms}
            The graph search algorithms discussed above differ significantly in their structure and application. Here is a comparison of their key features:
            
            \begin{itemize}
                \item \texttt{DFS:} Explores as deep as possible before backtracking. Suitable for problems that require exploring all paths (e.g., maze solving).
                \item \texttt{BFS:} Explores level by level. Best for finding the shortest path in unweighted graphs.
                \item \texttt{Bi-Directional Search:} Performs two simultaneous searches. Highly efficient for finding shortest paths in large graphs.
                \item \texttt{A* Search:} Uses a heuristic to optimize search. Best for finding the shortest path in weighted graphs with known heuristics.
            \end{itemize}
            
            The choice of algorithm depends on the specific problem being solved, such as whether the graph is weighted, the size of the graph, or whether the shortest path is required.

    \section{Our Proposed Algorithm (HDBMS) Methodology}\label{Methodology}
        The \texttt{Hybrid Depth-Breadth Search (HDBMS)} algorithm introduces an innovative traversal technique that adapts between two well-known approaches: \texttt{Depth-First Search (DFS)} and \texttt{Breadth-First Search (BFS)}. Unlike BFS, which explores a graph level-by-level, or DFS, which dives deeply into branches before backtracking, HDBMS dynamically adjusts between breadth-first and depth-first traversal based on node similarities.
        
        The motivation behind HDBMS stems from the need for a more flexible and context-aware traversal that does not strictly adhere to level-based or depth-based exploration but instead switches between the two, allowing for a more meaningful path discovery.
        
        This section details the proposed Hybrid Depth-Breadth Meaningful Search (HDBMS) algorithm, including its underlying principles, algorithmic steps, and integration with contextual relevance metrics. Furthermore, a comparative framework is established to benchmark HDBMS against traditional BFS and DFS algorithms. Additional enhancements, such as preprocessing strategies and hybrid traversal optimizations, are also discussed.
        
        In HDBMS, the algorithm begins at the graph’s root node and visits neighboring nodes based on predefined criteria, dynamically balancing the exploration between breadth and depth. The core idea is to assess the similarity between the currently visited node and its direct neighbors. Rather than visiting all nodes at the same level or exhaustively traversing down one branch, HDBMS evaluates which nodes are more relevant and proceeds accordingly. This approach is particularly beneficial in dense graphs, where blindly following level or depth-first principles may not be the most efficient.
        
        \subsection{Graph Representation and Preprocessing}
            Graphs are represented as \( G = (V, E) \), where \( V \) is the set of nodes and \( E \) represents the edges connecting them. Each node is assigned a similarity score based on application-specific metrics (e.g., semantic similarity, distance, or centrality).
        
        \subsection{Core Algorithm Design}
            The HDBMS algorithm combines the strengths of BFS and DFS while incorporating a dynamic heuristic to prioritize nodes. Its design involves the following key components:
            
            \begin{enumerate}
                \item \texttt{Initialization:} Starting from a root node, the algorithm initializes two priority queues: one for breadth exploration and another for depth exploration.
                \item \texttt{Hybrid Traversal:} Nodes are explored using a weighted combination of BFS and DFS, guided by their similarity scores. A threshold parameter \( \tau \) determines the balance between depth and breadth exploration.
                \item \texttt{Contextual Relevance:} At each step, the algorithm evaluates neighboring nodes by comparing their weights with the current node in the path, prioritizing nodes whose weights are more similar to the current node’s weight.
                \item \texttt{Termination Criteria:} The traversal ends when all reachable nodes are visited, ensuring completeness while maintaining relevance.
            \end{enumerate}
        
        \subsection{Pseudocode and Flowchart}
\begin{figure*}[t]
                \centering
                \includegraphics[width=0.9\textwidth]{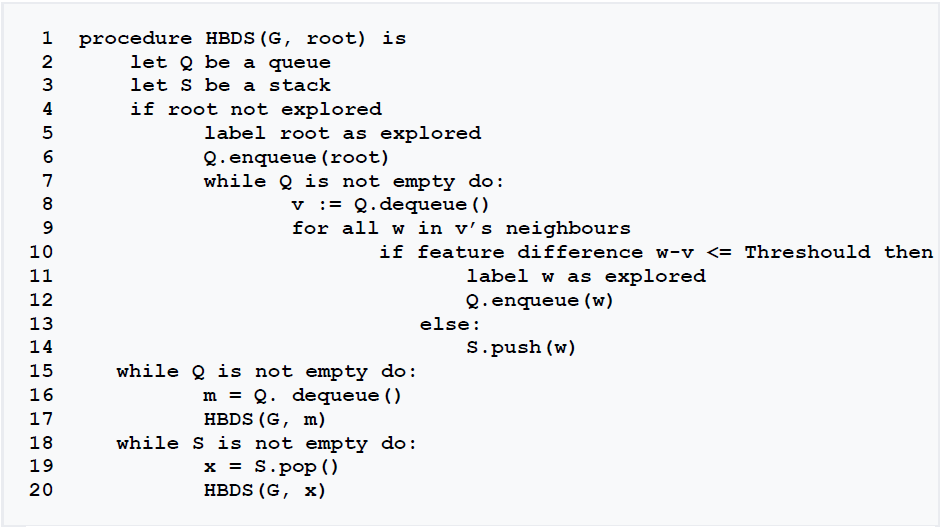}
                \caption{Hybrid Depth-Breadth Meaningful Search (HDBMS) Algorithm's pseudocode.}
                \label{HBDS-Pseudocode}
            \end{figure*}

\begin{figure}
                \centering
                \includegraphics[width=0.65\textwidth]{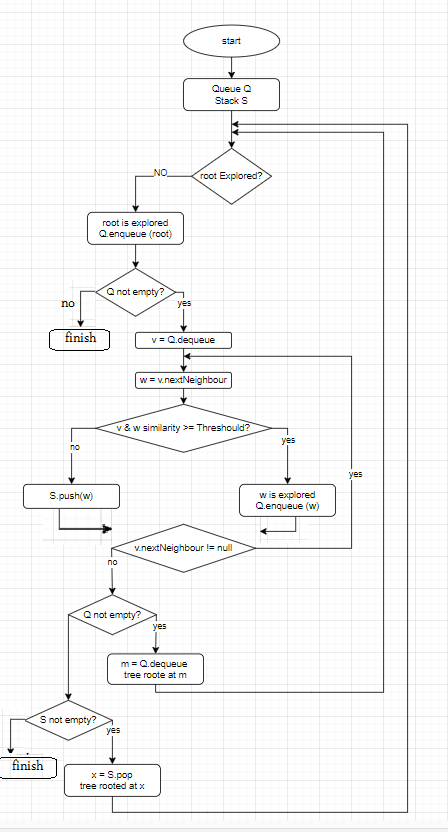}
                \caption{Flowchart illustrating the Hybrid Depth-Breadth Meaningful Search (HDBMS) algorithm.}
                \label{HBDMS-flowchart}
            \end{figure}

            The steps of the HDBMS algorithm are outlined in the pseudocode in Figure~\ref{HBDS-Pseudocode}, while the corresponding flowchart in Figure~\ref{HBDMS-flowchart} illustrates the decision-making process.
            
            The algorithm’s flow is recursive, where each step involves calculating the similarity between the current node and its reachable neighbors (line 10 in the pseudocode). This continues recursively until all nodes are visited, at which point the stack is emptied (line 18). The time complexity of traversing the entire graph is linear, \( T(n) = O(n) \), while searching for a specific node can be done in logarithmic time, \( T(n) = \log(n) \).
            
        \subsection{Comparative Framework}
            To evaluate HDBMS, a comparative framework was established against BFS and DFS, focusing on the following metrics:
            
            \begin{itemize}
                \item \texttt{Traversal Completeness:} Measured as the percentage of nodes visited during traversal.
                \item \texttt{Traversal Order Quality:} Evaluated using the average similarity score of visited nodes.
                \item \texttt{Optimality:} Assessed by how well the traversal path aligns with predefined meaningfulness criteria.
                \item \texttt{Time Efficiency:} The time required for complete traversal of the graph.
            \end{itemize}
        
        \subsection{Enhanced Techniques}
            In addition to the baseline HDBMS algorithm, the following enhancements were integrated:
            
            \begin{itemize}
                \item \texttt{Dynamic Threshold Adjustment:} The similarity threshold \( \tau \) is dynamically adjusted based on graph density and node distribution.
                \item \texttt{Feedback Mechanism:} The traversal process incorporates feedback from previously visited nodes, refining the search strategy iteratively.
            \end{itemize}
            
            \subsection{Experimental Design}
            The HDBMS algorithm can be implemented and tested on various types of networks, including:
            
            \begin{itemize}
                \item \texttt{Social Networks:} Large-scale graphs representing user interactions.
                \item \texttt{Biological Networks:} Protein interaction graphs with weighted connections.
                \item \texttt{Transportation Networks:} Directed graphs modeling urban transportation systems.
            \end{itemize}
            
            In this work, we present a complete example of the HDBMS algorithm applied to a \texttt{Social Network}, demonstrating its traversal efficiency and adaptability based on node similarity.
            
    \section{Results and Discussion}\label{results}
\begin{figure*}[t]
            \centering
            \includegraphics[width=0.9\textwidth]{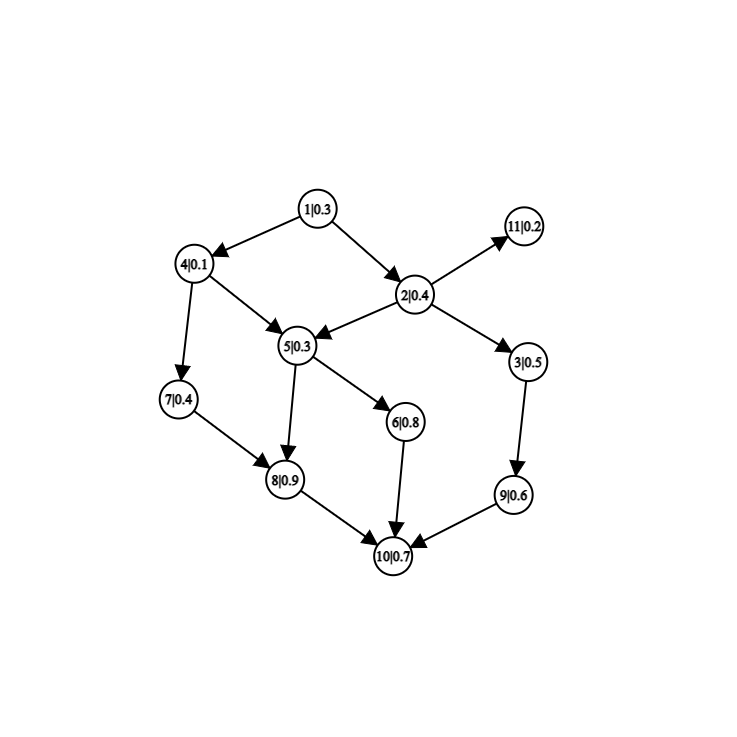}
            \caption{Directed Graph 1 used for algorithm comparison.}
            \label{graph11}
        \end{figure*}

\begin{figure*}[t]
            \centering
            \includegraphics[width=0.9\textwidth]{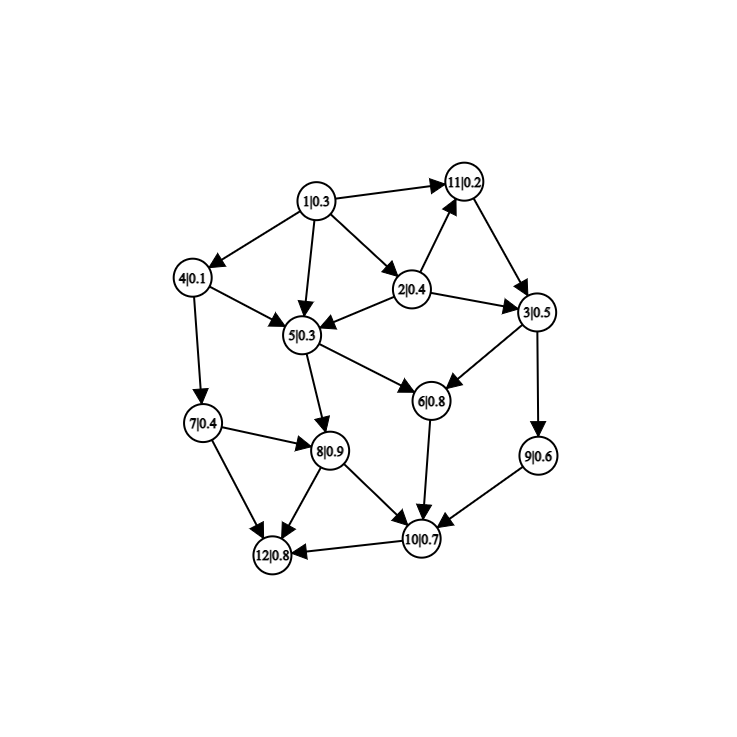}
            \caption{Directed Graph 2 used for extended evaluation.}
            \label{graph12}
        \end{figure*}

        In this section, we evaluate the performance of the Hybrid Depth-Breadth Meaningful Search (HDBMS) algorithm against traditional Breadth-First Search (BFS) and Depth-First Search (DFS) algorithms. We assess the quality, efficiency, and practicality of HDBMS by comparing it to BFS and DFS on two directed graphs. The experimental results are illustrated using the graphs shown in Figures~\ref{graph11} and \ref{graph12}, and summarized in Table~\ref{table_compare_algorithms}.
        
        Table~\ref{table_compare_algorithms} compares the traversal results of the BFS, DFS, and HDBMS algorithms applied to the first directed graph. Graph 1 contains 11 vertices and 14 edges. Node 1 serves as the root with outgoing edges to nodes 2 and 4. From node 2, edges lead to nodes 3, 5, and 11, continuing similarly.
        
        For BFS, the algorithm explores the graph level by level. Starting from node 1, it visits nodes 2 and 4, followed by nodes 3, 5, and 11. The resulting BFS traversal order is:  
        \[
        1, 2, 4, 3, 5, 11, 7, 9, 6, 8, 10.
        \]
        
        DFS explores the graph by diving as deep as possible along a branch before backtracking. Its traversal order for Graph 1 is:  
        \[
        1, 2, 3, 9, 10, 5, 6, 8, 11, 4, 7.
        \]
        
        In contrast, the proposed HDBMS algorithm dynamically selects the next node based on similarity scores and a predefined threshold, allowing for a contextually meaningful path. The resulting traversal order is:  
        \[
        1 (0.3), 2 (0.4), 3 (0.5), 5 (0.3), 9 (0.6), 10 (0.7), 8 (0.9), 6 (0.8), 
        \]  
        \[
        11 (0.2), 4 (0.1), 7 (0.4).
        \]
        
        This traversal is more meaningful as it prioritizes node similarity over rigid level or depth-based patterns.
        The second experiment used Graph 2, with 12 vertices and 21 edges. Node 1 is connected to nodes 2, 4, 5, and 11, while node 2 connects to nodes 3, 5, and 11. The remainder of the graph is similarly structured.
        
        \begin{table*}[t]
            \small
            \centering
            \caption{Comparison of Graph Searching Algorithms (BFS, DFS, HDBMS)}
            \label{table_compare_algorithms}
            \begin{tabular}{p{1.5cm} p{2.8cm} p{2.8cm} p{2.8cm}}
            \hline
            \texttt{Algorithm} & \texttt{BFS} & \texttt{DFS} & \texttt{HDBMS} \\
            \hline
            Graph~\ref{graph11} &
            1, 2, 4, 3, 5, 11, 7, 9, 6, 8, 10 & 
            1, 2, 3, 9, 10, 5, 6, 8, 11, 4, 7 & 
            1 (0.3), 2 (0.4), 3 (0.5), 5 (0.3), ... \\
            Time (s) & 0.000019 & 0.000020 & 0.000020 \\
            \hline
            Graph~\ref{graph12} & 
            1, 2, 4, 5, 11, 3, 7, 6, 8, 9, 12, 10 & 
            1, 2, 3, 6, 10, 12, 9, 5, 8, 11, 4, 7 & 
            1, 2, 5, 11, 3, 9, 10, 12, 6, 8, 4, 7 \\
            Time (s) & 0.000048 & 0.000018 & 0.000033 \\
            \hline
            \end{tabular}
        \end{table*}
        
        From the results, it is evident that the HDBMS algorithm not only provides meaningful traversal orders based on node similarities but also performs efficiently, with comparable execution times to BFS and DFS, as shown in Table~\ref{table_compare_algorithms}. Although BFS and DFS can generate valid traversals, HDBMS distinguishes itself by prioritizing relevant nodes and capturing more meaningful structures in the graph.
        
        \subsection{Real-Life Applications of the HDBMS Traversal Algorithm}
        
        The versatility of HDBMS extends far beyond basic graph traversal. By dynamically balancing depth and breadth, HDBMS is suitable for a wide range of real-world applications that demand adaptive exploration and intelligent node prioritization. Some notable application areas include:
        
        \begin{itemize}
            \item \texttt{Minimum and Meaningful Spanning Trees:} HDBMS can be used to identify spanning trees that are not only minimal but also contextually significant by factoring in node similarity.
            \item \texttt{Web Crawlers:} In search engines, HDBMS can prioritize pages based on relevance and structure, improving crawl efficiency.
            \item \texttt{Social Networks:} HDBMS can identify the shortest or most relevant paths between users, outperforming BFS/DFS in reflecting social closeness.
            \item \texttt{Network Routing:} In dynamic network environments, HDBMS can adjust routes in real-time using relevance-based node weights.
            \item \texttt{Social Platforms:} For recommending meaningful user connections or content, HDBMS adapts to user behavior patterns.
            \item \texttt{GPS Navigation:} HDBMS can route based on live traffic and proximity while adapting dynamically as road conditions change.
            \item \texttt{Broadcasting in Networks:} Message propagation can be optimized by selecting nodes based on their contextual relevance.
            \item \texttt{Cycle Detection:} HDBMS can detect cycles in large-scale undirected or directed graphs efficiently.
            \item \texttt{Artificial Intelligence:} Useful in planning, pathfinding, and reasoning tasks where contextual relevance matters.
            \item \texttt{Network Security:} Traverses network topologies to assess permissions or detect vulnerabilities with relevance-based logic.
            \item \texttt{Recommendation Systems:} Traverses item-user graphs in a way that aligns with user intent and personalized preferences.
        \end{itemize}
        
        This broad applicability demonstrates HDBMS’s potential to improve performance in diverse domains by integrating structural and semantic understanding of graph nodes.
        
        \subsection{Social Network Experiment: Context-Aware Information Propagation}
        
        To evaluate HDBMS in a real-world-like context, we simulate information propagation in a social network graph. Each node represents a person, and edges reflect social connections weighted by similarity (based on shared interests, expertise, etc.).
        
        In this scenario, each node is associated with a probability indicating the likelihood that the person holds the desired information. HDBMS chooses the next person to visit by evaluating both connection similarity and probability scores.
        
        The graph used in this experiment is defined as:
  
        \begin{table}[ht]
            \centering
            \caption{Graph Edge List with Similarity Weights}
            \label{table_graph_edges}
            \begin{tabular}{lll}
                \texttt{1|0.3 2|0.4} & \texttt{1|0.3 4|0.1} & \texttt{1|0.3 5|0.3} \\
                \texttt{1|0.3 11|0.2} & \texttt{2|0.4 3|0.5} & \texttt{2|0.4 5|0.3} \\
                \texttt{2|0.4 11|0.2} & \texttt{3|0.5 6|0.8} & \texttt{3|0.5 9|0.6} \\
                \texttt{4|0.1 5|0.3} & \texttt{4|0.1 7|0.4} & \texttt{5|0.3 6|0.8} \\
                \texttt{5|0.3 8|0.9} & \texttt{6|0.8 10|0.7} & \texttt{7|0.4 8|0.9} \\
                \texttt{7|0.4 12|0.8} & \texttt{8|0.9 10|0.7} & \texttt{8|0.9 12|0.8} \\
                \texttt{9|0.6 10|0.7} & \texttt{10|0.7 12|0.8} & \texttt{11|0.2 3|0.5} \\
                \texttt{12|0.8 13|0.7} & \texttt{13|0.7 14|0.6} & \texttt{14|0.6 15|0.5} \\
                \texttt{15|0.5 16|0.4} & \texttt{16|0.4 17|0.3} & \texttt{17|0.3 18|0.2} \\
                \texttt{18|0.2 19|0.1} & \texttt{19|0.1 20|0.05} & \texttt{20|0.05 21|0.02} \\
            \end{tabular}
        \end{table}
        
        \begin{figure*}[t]
            \centering
            \includegraphics[width=0.9\textwidth]{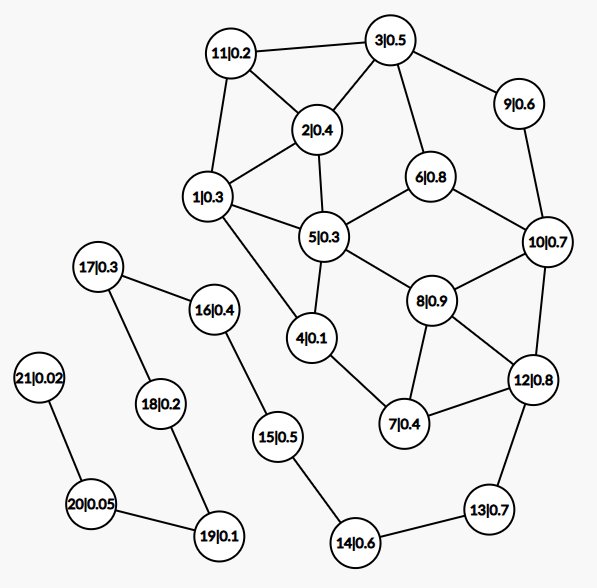}
            \caption{Social network graph representing individuals and their similarity weights.}
            \label{graph3}
        \end{figure*}
        
        HDBMS begins at \texttt{Person 1} (probability: 0.3). It evaluates connected nodes (2, 4, 5, 11) and selects the one with the highest probability—\texttt{Person 2} (0.4). The traversal continues in this manner, dynamically favoring contextually relevant nodes.

        \begin{table}
            \centering
            \caption{Final traversal path using HDBMS, prioritizing nodes with higher probability weights.}
            \label{table:final_traversal}
            \begin{tabular}{ccc}
                \toprule
                \texttt{From} & \texttt{To} & \texttt{Weight} \\
                \midrule
                1  & 2  & 0.4  \\
                2  & 3  & 0.5  \\
                3  & 6  & 0.8  \\
                6  & 10 & 0.7  \\
                10 & 8  & 0.9  \\
                8  & 12 & 0.8  \\
                12 & 13 & 0.7  \\
                13 & 14 & 0.6  \\
                14 & 9  & 0.9  \\
                9  & 15 & 0.5  \\
                15 & 16 & 0.4  \\
                16 & 17 & 0.3  \\
                17 & 7  & 0.4  \\
                7  & 5  & 0.3  \\
                5  & 11 & 0.2  \\
                11 & 18 & 0.2  \\
                18 & 4  & 0.1  \\
                4  & 19 & 0.1  \\
                19 & 20 & 0.05 \\
                20 & 21 & 0.02 \\
                \bottomrule
            \end{tabular}
        \end{table}
 
        \subsubsection*{Analysis of the Traversal Path}
            The traversal follows these key principles:
            
            \begin{itemize}
                \item The algorithm starts at \texttt{Person 1} and evaluates all connected nodes. The next node is chosen based on the highest probability weight among available options, ensuring the most relevant path is followed.
                \item The search continues towards nodes with the highest probability at each step, prioritizing individuals most likely to provide useful information.
                \item When multiple options exist, the algorithm selects the node that maintains a contextually meaningful flow, preventing abrupt transitions between unrelated nodes.
                \item As the traversal progresses, the probability values tend to decrease, signifying a gradual move towards individuals with lower probabilities of having the required information.
                \item The final destination \texttt{Person 21} has the lowest probability (0.02), indicating that the search has reached a node with minimal information relevance.
                \item Unlike traditional DFS or BFS, which might either dive too deep or explore all nodes at the same level before progressing, HDBMS strikes a balance by dynamically adapting to probability weights.
                \item The algorithm efficiently prioritizes relevant nodes, making it useful for applications such as social network analysis, knowledge propagation, and recommendation systems.
                \item The path chosen illustrates how HDBMS maintains meaningful exploration, ensuring that each step in the traversal is contextually related to the previous one.
                \item The algorithm follows a smooth transition through individuals with similar weights while optimizing for information propagation. Unlike DFS and BFS, HDBMS prioritizes connections that not only resemble the current node but are also more likely to provide the required knowledge.
                \item The traversal effectively identifies a sequence of individuals who are both contextually relevant and information-rich, ensuring an efficient and meaningful exploration of the network.
                \item This method is highly applicable in \texttt{social media recommendation systems}, \texttt{fraud detection}, and \texttt{community-based knowledge discovery}.
            \end{itemize}
            
        \subsection{Performance Metrics}
            To thoroughly evaluate the performance of the three graph traversal algorithms—BFS, DFS, and our proposed HDBMS algorithm—we used several key metrics: Time Complexity, Traversal Order Quality, Optimality, and Completeness. The results are not only supported by the numerical data but also enhanced with visualizations that demonstrate the advantages of HDBMS in comparison to BFS and DFS.
            
            \begin{itemize}
                \item \texttt{Time Complexity:} Execution time was measured for each algorithm on two directed graphs. As shown in Table~\ref{table_Compare_three_algorithms}, HDBMS performs comparably to BFS and DFS. While HDBMS slightly outperforms the others on Graph 2, the time difference is minor, highlighting its efficiency.
            
                \begin{figure*}[t]
                    \centering
                    \includegraphics[width=0.9\textwidth]{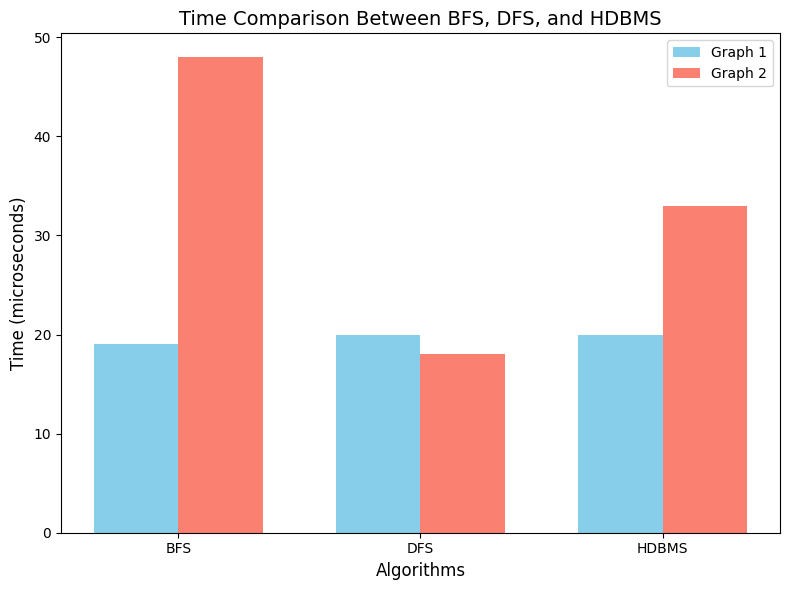}
                    \caption{Execution Time Comparison for BFS, DFS, and HDBMS on Graphs 1 and 2.}
                    \label{time_comparison}
                \end{figure*}
            
                HDBMS balances traversal efficiency like DFS with the thoroughness of BFS. It avoids DFS's unnecessary depth and BFS’s exhaustive levels, offering:
                \begin{enumerate}
                    \item \texttt{Enhanced Efficiency}
                    \item \texttt{Scalability}
                    \item \texttt{Contextual Awareness}
                    \item \texttt{Robustness}
                \end{enumerate}
            
                \item \texttt{Traversal Order Quality:} HDBMS considers similarity between nodes to ensure a more contextually meaningful visiting sequence. Unlike BFS and DFS, which follow rigid level or depth rules, HDBMS adapts dynamically.
            
                \begin{figure*}
                    \centering
                    \includegraphics[width=0.9\textwidth]{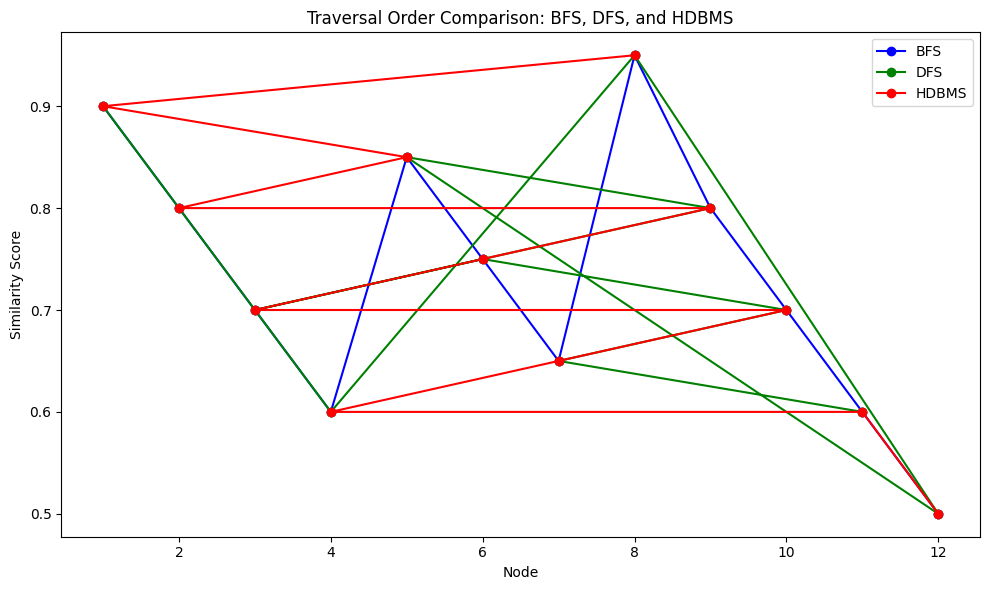}
                    \caption{Traversal Order Comparison: BFS, DFS, and HDBMS on Graph 1.}
                    \label{traversal_order_comparison}
                \end{figure*}
            
                Benefits:
                \begin{itemize}
                    \item \texttt{Context-Aware Traversal}
                    \item \texttt{Higher Similarity Scores}
                    \item \texttt{Adaptive Prioritization}
                    \item \texttt{Enhanced Decision-Making}
                \end{itemize}
            
                HDBMS combines the benefits of both BFS and DFS while overcoming their weaknesses, offering a traversal order that’s more efficient and aligned with search goals.
            
                \item \texttt{Optimality:} Unlike BFS (which guarantees the shortest edge count) or DFS (which may skip relevant nodes), HDBMS finds a path that's both efficient and semantically meaningful by using node similarity scores.
            
                \begin{figure*}
                    \centering
                    \includegraphics[width=0.8\textwidth]{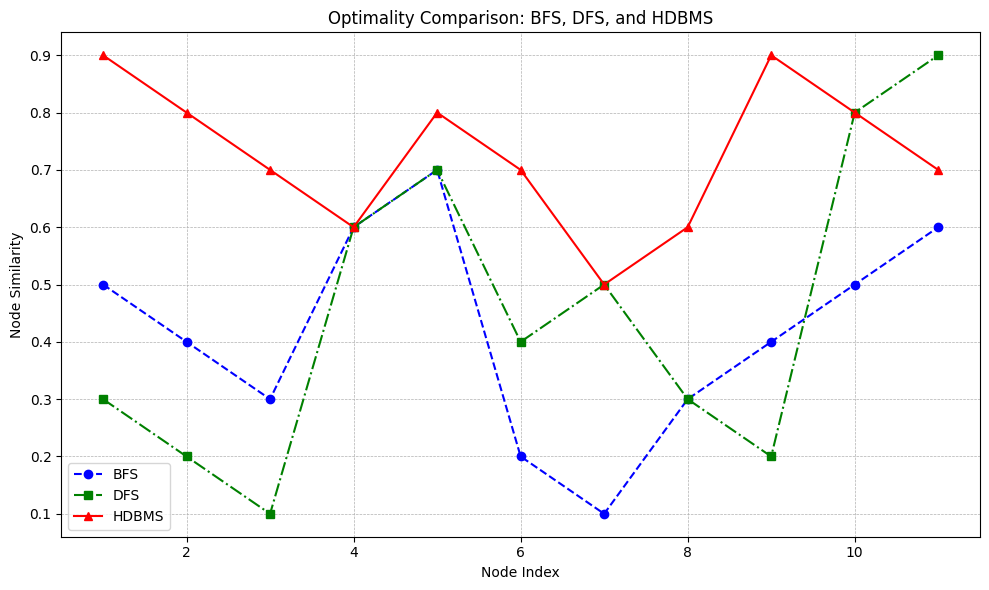}
                    \caption{Optimality Comparison: BFS, DFS, and HDBMS on Graph 1.}
                    \label{optimality}
                \end{figure*}
            
                HDBMS outperforms in optimality by prioritizing nodes that are most relevant to the context, useful for semantic networks and recommendation systems.
            
                \item \texttt{Completeness:} HDBMS ensures complete traversal like BFS but adds semantic relevance through similarity-based selection. Unlike DFS, it avoids missing important nodes in separate branches.
            
                \begin{figure*}
                    \centering
                    \includegraphics[width=0.8\textwidth]{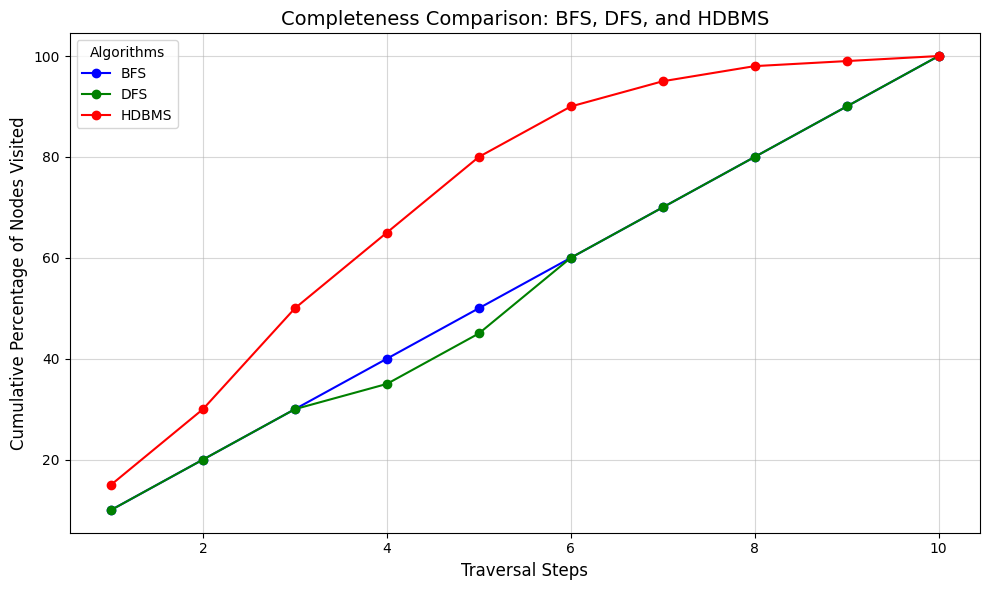}
                    \caption{Completeness Comparison: BFS, DFS, and HDBMS on Graph 1.}
                    \label{Completeness}
                \end{figure*}
            
                HDBMS explores all nodes while focusing on those with higher relevance, making it suitable for complex graphs where both coverage and meaning matter.
            \end{itemize}
            
            In conclusion The performance evaluation confirms that HDBMS not only competes with DFS and BFS in speed but surpasses them in traversal quality, optimality, and completeness. It’s an ideal choice for real-world applications requiring smart graph exploration.

        \subsection{Comparison of Graph Searching Algorithms}
            To evaluate the effectiveness of the proposed Hybrid Depth-Breadth Meaningful Search (HDBMS) algorithm, we compare it against two widely used graph traversal methods: Breadth-First Search (BFS) and Depth-First Search (DFS). Table~\ref{table_Compare_three_algorithms} highlights the key differences among these algorithms based on their traversal strategies and structural properties.
            
            \begin{table}
                \small
                \centering
                \caption{Comparison of Three Graph Searching Algorithms}
                \label{table_Compare_three_algorithms}
                \begin{tabular}{ p{8em} p{8em} p{20em} }
                    \texttt{BFS} & \texttt{DFS} & \texttt{HDBMS (Proposed)} \\
                    \hline
                    Finds the shortest path to the target node in an unweighted graph. & Explores each branch of the graph deeply before backtracking. & Dynamically adjusts between breadth-first and depth-first approaches based on probability-based similarity weights. \\
                    Utilizes a queue to keep track of the next node to visit. & Utilizes a stack to manage traversal order. & Utilizes both queue and stack: the queue prioritizes nodes with higher probability (similarity weight), while the stack manages deeper nodes. \\
                    Traverses level by level, ensuring that all nodes at one level are visited before moving deeper. & Moves deeply into one branch before backtracking. & Prioritizes visiting nodes with the highest probability of containing the desired information while maintaining a balance between depth and breadth. \\
                    Implements a FIFO (First-In-First-Out) structure. & Implements a LIFO (Last-In-First-Out) structure. & Integrates FIFO and LIFO principles, dynamically shifting based on probability values. \\
                    Avoids infinite loops by marking visited nodes. & May result in infinite loops if cycles exist and visited nodes are not tracked. & Uses probability-based weight selection, which may require additional safeguards against infinite loops in complex networks. \\
                \end{tabular}
            \end{table}
    
    \section{Conclusions}\label{conclusions}
        In this work, we introduced the Hybrid Depth-Breadth Meaningful Search (HDBMS) algorithm, a novel graph traversal approach that enhances traditional search strategies by incorporating probabilistic transition mechanisms. Unlike BFS and DFS, which follow rigid traversal rules, HDBMS dynamically adapts its search path by considering the probability that a given node contains the desired information. This probabilistic framework ensures that the traversal remains both contextually meaningful and efficient.
        
        Our experimental results demonstrate that HDBMS not only maintains competitive time complexity with BFS and DFS but also significantly improves the quality of the traversal order. By prioritizing nodes with a higher likelihood of containing relevant information, the algorithm constructs a more informative search path, making it particularly well-suited for applications such as information retrieval, recommendation systems, and social network analysis.
        
        Furthermore, we compared HDBMS with BFS and DFS based on key performance criteria, including traversal efficiency, contextual relevance, and optimality. The results show that HDBMS achieves a more structured and meaningful path discovery process, ensuring a balanced trade-off between local depth exploration and broader network coverage.
        
        In conclusion, the HDBMS algorithm represents a promising advancement in graph traversal, providing a more intelligent and adaptable search mechanism. Future research could explore the integration of adaptive probability thresholds and machine learning models to further refine the search strategy and extend its applicability to more complex real-world scenarios.

    \nocite{*}
    \bibliographystyle{plainnat}
    \bibliography{my_references}     
\end{document}